\documentclass[global,twocolumn]{svjour230703}

\date{23.07.2003}
\usepackage{graphicx}

\title{Dipolar Relaxation in an ultra-cold Gas of magnetically trapped chromium atoms}

\author{S. Hensler\inst{1}\and J. Werner\inst{1}\and A. Griesmaier\inst{1}\and P.O. Schmidt\inst{1}\and A. G\"{o}rlitz\inst{1}\and T.
Pfau\inst{1}\and \\S. Giovanazzi\inst{2}\and
\\K. Rz\c{a}\.{z}ewski\inst{3}}

\institute{5. Physikalisches Institut, Universit\"{a}t Stuttgart,
Pfaffenwaldring 57, D-70550 Stuttgart, Germany\and School of
Physics and Astronomy, University of St Andrews, North Haugh, St
Andrews, Fife, KY 16 9SS, Scotland\and Center for Theoretical
Physics and College of Science, Polish Academy of Science, Aleja
Lotnik\'{o}w 32/46, 02-668 Warsaw, Poland}

\begin{document}

\maketitle

\begin{abstract}
We have investigated both theoretically and experimentally dipolar
relaxation in a gas of magnetically trapped chromium atoms. We
have found that the large magnetic moment of 6\,$\mu_B$ results in
an event rate coefficient for dipolar relaxation processes of up
to $3.2\cdot10^{-11}$\,cm$^{3}$s$^{-1}$ at a magnetic field of
44\,G. We present a theoretical model based on pure dipolar
coupling, which predicts dipolar relaxation rates in agreement
with our experimental observations. This very general approach can
be applied to a large variety of dipolar gases.
\end{abstract}

\section{Introduction}

\label{sec:intro}

Recently, the dipole-dipole interaction in Bose-Einstein
condensates has generated significant theoretical interest
\cite{Baranov2002}. It has been shown that the long-range
character and the anisotropic nature of the interaction between
magnetic or electric dipoles will lead to many interesting new
phenomena like novel quantum phase transitions \cite{Goral2002},
dipolar order \cite{Pu2001} and spin tunnelling \cite{Pu2002} in
the condensate. New questions concerning stability
\cite{Goral2000,Yi2001} and tunability of the interaction
\cite{Giovanazzi2002} of such a condensate arise. In most of the
recent work, the interaction between dipoles is treated as
completely elastic. However, in a typical experiment aimed at the
creation of a Bose-Einstein condensate, the atoms are held in a
magnetic trap in the energetically highest magnetic sub-level.
Therefore, inelastic dipolar relaxation processes in which the
atomic spin flips and the Zeeman energy is released cannot
\textit{a priori} be neglected.

Most of the above proposals require a large dipole moment like the
one of a polar molecule in an electric field. In this paper, we
show theoretically and experimentally that already for chromium
with a magnetic moment of 6\,$\mu_B$ heating and atom loss due to
dipolar relaxation is so significant that it is impossible to
reach Bose-Einstein condensation by the standard procedure of
rf-evaporation in a magnetic trap.

Using a simple model which only takes into account the dipolar
interaction and neglects all other aspects of the molecular
interaction on the scattering process we are able to explain the
observed dipolar relaxation rate in ultra-cold $^{52}$Cr.
Qualitatively, we find the same behavior for $^{50}$Cr. This gives
strong evidence that, in the absence of resonant behavior,
inelastic collisions due to dipolar relaxation can be treated in a
very general way.

\section{Dipolar Relaxation: Theory}

\label{sec:theory}

As a consequence of Maxwell's equations it is impossible to create
a maximum of a static magnetic field in a region where external
currents are absent. For this reason atoms are usually
magnetically trapped in the energetically highest Zeeman sub-level
(weak-field-seeking state). The trap lifetime of these atoms is
mainly limited by collisions. Aside from background gas collisions
and three-body recombination, the atoms undergo spin-exchange and
dipolar relaxation collisions. In contrast to spin-exchange
collisions, where the total spin is conserved, the spin is flipped
in a dipolar relaxation process since dipole-dipole interaction
does not preserve the total spin, but rather the total angular
momenta. Therefore, if one considers a cloud of fully polarized
atoms in the energetically highest state, only spin relaxation
which does not conserve the total spin leads to atom loss from the
magnetic trap.

Here, we estimate the inelastic cross sections for spin-flip
transitions due to the magnetic spin-dipole spin-dipole
interaction using Fermi's Golden rule. A similar form of this
approximation, equivalent to the Born approximation, has already
been considered in \cite{Shlyapnikov1994,Fedichev1996}, for
calculating the dipolar relaxation rates of meta-stable triplet
helium. In particular, in \cite{Fedichev1996} it is shown that the
bare Born approximation is reasonably consistent with the
distorted-wave approximation which is a true first order
calculation. Because of the small dipole-dipole coupling in
polarized triplet helium ($S=1$), the latter is considered to be
an almost exact approach.

For chromium, the situation is slightly different because the
magnetic coupling is larger. A first order calculation, in the
form of the Born approximation, provides us with an approximate
analytical tool for the interpretation of the experimental
results, but may not be expected to be as accurate as in the case
of metastable helium.

We consider for simplicity the case of two identical atoms trapped
in the energetically highest sub-level of the Zeeman manifold.
Moreover, we consider the case of zero electronic angular momenta
$L=0$ such that the magnetic interaction solely arises from the
electronic spin. The long range part of the interaction between
two atomic magnetic dipole moments $\vec{\mu}_1=g_S\mu_B\vec{S}_1$
and $\vec{\mu}_2=g_S\mu_B\vec{S}_2$ located at ${\bf r}_1$ and
${\bf r}_2$, where $\vec{S}_1$ and $\vec{S}_2$ are the two spin
matrices, reads
\begin{eqnarray}
U_{\mathrm{dd}}(\vec{r}) =  \mu_0(g_S\mu_B)^2 \, {(\vec{S}_1 \cdot
\vec{S}_2) - 3\, (\vec{S}_1 \cdot \hat{\vec{r}}) \, (\vec{S}_2
\cdot \hat{{\bf r}})  \over 4 \pi r^3}. \label{interaction0}
\end{eqnarray}
Here we have introduced the interatomic separation $\vec{r} =
\vec{r}_2 - \vec{r}_1$ with $\hat{\vec{r}} = \vec{r}/r$ and the
magnetic permeability of the vacuum $\mu_0$. The tensorial part of
the dipolar interaction (\ref{interaction0}), namely $(\vec{S}_1
\cdot \vec{S}_2) - 3\, (\vec{S}_1 \cdot \hat{\vec{r}}) \,
(\vec{S}_2 \cdot \hat{\vec{r}}) $ can be rewritten in terms of
spin-flip operators:
\begin{eqnarray}\label{tensore}
S_{1z} \cdot S_{2z} + {1\over 2} \left(S_{1+} \cdot S_{2-} +
S_{1-} \cdot S_{2+}\right)\nonumber\\
- {3\over4} \left(2\;\hat{z} S_{1z} + \hat{r}_- S_{1+} + \hat{r}_+
S_{1-}\right)\\ \times \left(2 \;\hat{z} S_{2z} + \hat{r}_- S_{2+}
+ \hat{r}_+ S_{2-}\right)\nonumber
\end{eqnarray}
where
\begin{eqnarray}
\hat{z}={z\over r} ,\;\;\;\;\; \hat{r}_+={x+iy\over r},\;\;\;\;\;
\hat{r}_-={x-iy\over r}, \label{angles}
\end{eqnarray}
and where $S_{+}=(S_{x} + i S_{y})/\sqrt{2}$ and $S_{-}=(S_{x} - i
S_{y})/\sqrt{2}$ are ladder operators of the spin operators. From
the above expression (\ref{tensore}) it can be seen that in first
order Born approximation each single atom can flip at most one
spin ($\Delta m_S=1$) and in total for the two-atoms system
$\Delta m_S=2$.

In the Born approximation the total cross section for two
identical colliding atoms in the same internal state can be
expressed in the following form
\begin{eqnarray}
\sigma = \left(\frac{ m}{4 \pi \hbar^2}\right)^2 {1 \over k k_f}
\left[ \int |\tilde{U}_{\mathrm{dd}}(\vec{k}-\vec{k}')|^2
\delta(|\vec{k}'|-k_f) \, d\vec{k}' \right. \nonumber\\
\left. \pm \int \tilde{U}_{\mathrm{dd}}(\vec{k}-\vec{k}')
\tilde{U}_{\mathrm{dd}}^*(-\vec{k}-\vec{k}')
\delta(|\vec{k}'|-k_f) \, d\vec{k}' \right] \, \label{sigmagen}
\end{eqnarray}
where $\vec{k}$ and $\vec{k}'$ are the initial and final wave
vectors, respectively, $k_f$ is the modulus of the final state
wave vector. $\tilde{U}_{\mathrm{dd}}(\vec{q})$ is the Fourier
transform of the dipole-dipole interaction (\ref{interaction0})
already contracted between the (symmetric) spin (or internal
state) functions of initial and final states. The sign $\pm$ of
the second term in the square bracket, the "exchange" term,
depends on particle's symmetry: $+$ holds for bosons and $-$ for
fermions. When the two colliding atoms are not in the same
internal state a generalization of (\ref{sigmagen}) with both
signs should be used.

\subsection{Dipolar transitions in a dipolar Bose gas}
We have in mind the specific case of $^{52}$Cr atoms in the
electronic ground state, magnetically trapped in the highest
Zeeman sub-level. For $^{52}$Cr, both the nuclear spin and the
electron orbital momentum vanish (I=0, L=0), and only the electron
spin is finite ($S=3$). The degeneracy of the 2S+1 atomic ground
state sub-levels is lifted by a magnetic field $B$. The Zeeman
splitting is given by
\begin{eqnarray}
\Delta E=g_S\mu_B B
\end{eqnarray}
where $\mu_B$ is the Bohr magneton and $g_S$ is the
Land$\acute{e}$ factor ($g_S=2$). The magnetic field is chosen
along the $z$ direction and determines the quantization axis. We
label the internal states via the $z$-component of the total
internal angular momentum (in this specific case only the
electronic spin), like $|m_S\rangle$.

\begin{figure}
    \begin{center}
    \includegraphics[width=0.8\columnwidth]{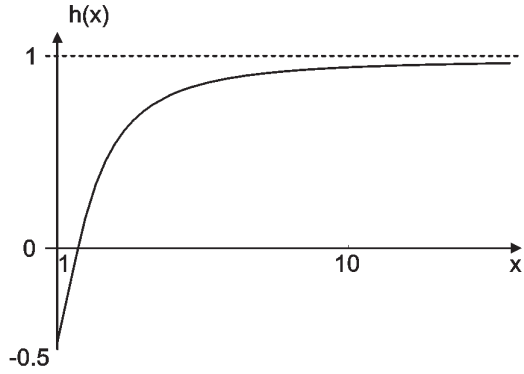}
    \end{center}
    \caption{Function h(x); h(x) is a monotonically increasing function
which asymptotically approaches 1.}\label{fig:hfunction}
\end{figure}

We consider specifically the case in which both atoms are prepared
in the stretched state, i.e $|m_S=s\rangle$. The total spin
function of the two colliding particles is already symmetric and
will be denoted as $|s;s\rangle$. The possible final internal
states of the two colliding atoms are
\begin{eqnarray}
|f_0\rangle&=&|s;s\rangle\\
|f_1\rangle&=&
{\left(|s;s-1\rangle+|s-1;s\rangle\right) \over \sqrt{2}}\\
|f_2\rangle&=&|s-1;s-1\rangle
\end{eqnarray}
where the subscripts $0,1,2$ refer to the number of spins that are
flipped. For dipolar scattering, the total cross section will
depend on the relative orientation of the initial relative
momentum ${\bf k}$ and the polarization axes. The averaged cross
sections over all possible orientations for the $0$, $1$ and $2$
spin-flip processes reads
\begin{eqnarray}
\sigma_0 &=&  {16\pi\over45} S^4 \left(\frac{\mu_0(g_S\mu_B)^2
m}{4
\pi \hbar^2}\right)^2 \left[1 + h(1)\right]\\
\sigma_1 &=&  {8\pi\over15} S^3 \left(\frac{\mu_0(g_S\mu_B)^2 m}{4
\pi \hbar^2}\right)^2 \left[1+h(k_f/k_i)\right]{k_f\over k_i}\\
\sigma_2 &=&  {8\pi\over15} S^2 \left(\frac{\mu_0(g_S\mu_B)^2 m}{4
\pi \hbar^2}\right)^2 \left[1+h(k_f/k_i)\right]{k_f\over k_i}.
\label{crosssections}
\end{eqnarray}
While $\sigma_1$ and $\sigma_2$ are the cross sections for dipolar
relaxation, $\sigma_0$ is the contribution to the elastic cross
section due to the dipole-dipole interaction. Here
\begin{eqnarray}
h(x)=-{1\over2}-{3\over8}{(1-x^2)^2\over
x(1+x^2)}\log\left({(1-x)^2\over (1+x)^2}\right)
\end{eqnarray}
represents the ratio of the exchange contribution to the direct
one of Eq. (\ref{sigmagen}). The function $h$ (see Fig
\ref{fig:hfunction}) is defined in the interval (1,$\infty$) and
exhibits a monotonic increase from h(1)=-1/2 to $h(\infty)=1$.
(For large values of the argument $h(x)\approx 1-4/x^2$.)

Note, that the term in brackets in Eq. (\ref{crosssections}) can
be expressed in terms of the classical electron radius $r_0$,
atomic mass $m$ and electronic mass $m_e$:
\begin{eqnarray}
\left(\frac{\mu_0(g_S \mu_B)^2 m}{4 \pi
\hbar^2}\right)^2=\left(\frac{m}{m_e}\right)^2 r_0^2 \,.
\label{r0}
\end{eqnarray}
One can see that the explicit dependence on $\hbar$ disappears.

The factor ${k_f \over k_i}$ accounts for the different density of
final states in the inelastic process ($\Delta E \neq 0$). The
ratio between final and initial wave-vectors is $k_f/k_i =
\sqrt{1+{m\Delta E \over \hbar^2 k_i^2}}$ for a one spin-flip
transition and $k_f/k_i = \sqrt{1+{2m\Delta E \over \hbar^2
k_i^2}}$ for a two spin-flip transition. If the Zeeman energy is
larger than the initial kinetic energy it exhibits a
characteristic $\sqrt{B}$ dependence. The expressions in Eq.
(\ref{crosssections}) for one and two spin flip processes can also
be applied to meta-stable helium ($S=1$). For $B=0$ we recover the
result of \cite{Shlyapnikov1994,Fedichev1996} for one and two
spin-flip process (using $ \langle v_{rel} \rangle = \sqrt{16
k_BT/\pi m}$ ). Our method yields an analytical solution for all
values of the magnetic field which reproduces the magnetic-field
dependence found in \cite{Shlyapnikov1994,Fedichev1996} at very
small and large magnetic field.

\section{Experiments}

\label{sec:experiments}

\subsection{Dipolar relaxation in chromium}

The magnetically trappable substates in the $^7S_3$ ground state
of chromium are the low-field seeking states $m_S=1,2,3$. In a
polarized sample of atoms in the energetically highest state
($m_S=3$), there are two possible decay channels due to dipolar
relaxation:
\begin{eqnarray}
    |3;3\rangle,|3;3\rangle & \rightarrow & |3;3\rangle,|3;2\rangle + \Delta E \nonumber \\
    |3;3\rangle,|3;3\rangle & \rightarrow & |3;2\rangle,|3;2\rangle + 2\Delta
    E.
\end{eqnarray}
In these processes either one or both colliding atoms change their
state to the $m_S=2$ state and release a Zeeman-energy of $\Delta
E$ or $2\Delta E$, respectively. The energy is equally distributed
among the colliding atoms. In a single spin flip transition the
increase in temperature which corresponds to the released Zeeman
energy $\left(3k_B\Delta T\right.=\left.\Delta E/2\right)$ is
250$\,\mu$K for an offset field of\linebreak $B_0$=$10$\,G. This
is generally not high enough to eject the atoms from the trap. As
a consequence thermalization leads to heating of the trapped cloud
(see Fig. \ref{fig:rfshield}(a)). The increasing population of the
$m_S=2$ state opens additional collision channels due to spin
exchange collisions and dipolar relaxation between identical and
different states. The sample becomes more and more depolarized and
only collisions with a final state of $m_S<1$ lead to atom loss.
In this paper, we concentrate on dipolar relaxation of $m_S=3$
state atoms in an almost spin-polarized sample.

\subsection{Methods to determine the dipolar relaxation rate}

In the following, we present three experimental methods from which
we obtain the dipolar relaxation coefficient $\beta_{dr}$ for the
$m_S=3$ state. It can be determined by measuring either the
decreasing population (method (i), (ii)) of this state or the
heating of the atom cloud in the trap (method (iii)) over time.
Methods (i) and (iii) have also been used in
\cite{Guery-Odelin1998} to determine the dipolar relaxation rate
of Cs in the F=3 ground state.

The atom loss from the $m_S=3$ state of an initially polarized
sample of $N_3$ atoms is described by
\begin{equation}
\frac{dN_3(t)}{dt}=-\int{\Gamma(\vec{r},t)n_3(\vec{r},t)\:d^3r},
\end{equation}
where $n_3(\vec{r} ,t)$ and $\Gamma(\vec{r},t)$ are the position
and time-dependent atom density and loss rate, respectively.
\linebreak$\Gamma(\vec{r},t)$ can be written as
\begin{equation}
\Gamma(\vec{r},t)=\gamma_{bg}+\beta_{dr}n_3(\vec{r},t)+\widetilde{\Gamma}(\vec{r},t).
\end{equation}
Here background gas collisions with a rate coefficient
$\gamma_{bg}$ are represented by the first term and the
density-dependent dipolar relaxation with a rate constant
$\beta_{dr}$ by the second term. Other collisional losses e.g.
three-body recombination, relaxation and spin-exchange processes
which involve states other than the $m_S=3$ state are included in
$\widetilde{\Gamma}(\vec{r},t)$. For low densities and small
populations in these other states, this term can be neglected.

\begin{figure}
\begin{minipage}{0.49\columnwidth}
    \includegraphics[width=1\columnwidth]{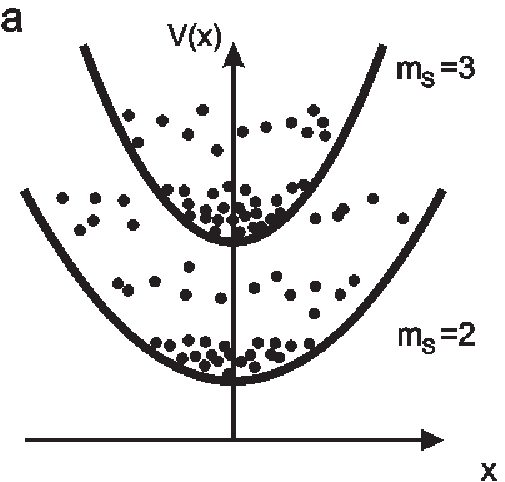}
\end{minipage}
\begin{minipage}{0.49\columnwidth}
    \includegraphics[width=1\columnwidth]{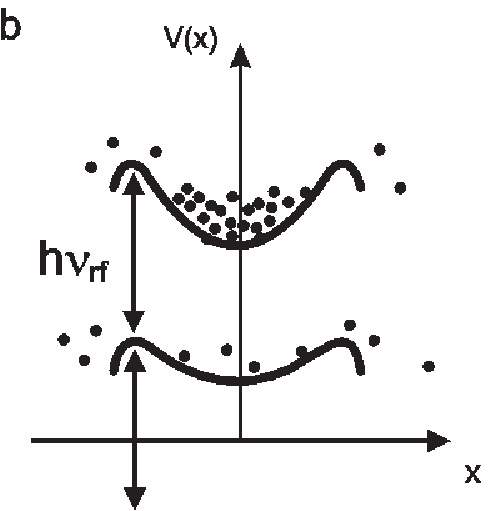}
\end{minipage}
    \caption{a) Dipolar relaxation in a magnetic trap. Atoms initially prepared in the $m_S=3$ state undergo dipolar relaxation and fill the $m_S=2$ state. Indicated is the energy gain of a colliding atom of $\Delta$E/2 and $\Delta$E of single and double spin flip transition, respectively. Thermalization of the released Zeeman energy leads to heating in the trap.   b) If a rf-shield is applied, the collisional products leave the trap. Thus, the collision process is turned into a real loss process.}\label{fig:rfshield}
\end{figure}

The first method (i) takes advantage of the Zeeman-energy gain
during a relaxation process. The energetic collisional products
are selectively removed by a radio-frequency (rf) shield with
frequency $\nu_{\mathit{rf}}$ as depicted in Fig.
\ref{fig:rfshield}. Thus, dipolar relaxation is turned into a real
loss process and can directly be measured by monitoring the
remaining atoms. The decay is fitted to the solution of the
differential equation:
\begin{equation}\label{eq:decayhighoffset}
    \frac{dN_3(t)}{dt}=-\gamma_{bg}
    N_3(t)-{\beta}^\circ_{dr}\frac{N_3^2(t)}{\overline{V}(t)}.
\end{equation}
Where $\overline{V}=\sqrt{8}(2\pi)^{3/2}\sigma_x\sigma_y\sigma_z$
is the mean volume of an atomic cloud with a Gaussian density
distribution with 1/$\sqrt{e}$ widths $\sigma_{x,y,z}$. To remove
the collisional products the rf-frequency of the shield has to be
chosen such that $m_J(h\nu_{\mathit{rf}}-\Delta E)<\Delta E/2$. In
this case $\beta^\circ_{dr}$ is an event rate which contains
single and double spin flip transitions. $\beta^\circ_{dr}$ can be
calculated from the thermal average over both inelastic collision
channels:
\begin{equation}
\label{eq:betahighoffset}
    \beta^\circ_{dr}=2\cdot\langle(\sigma_1+\sigma_2)v_{rel}\rangle_{_{therm}}.
\end{equation}
The factor of $2$ accounts for two removed atoms per collision
event. At the same time, the rf-frequency has to be adjusted to a
sufficiently high value, such that the initial $m_S=3$ cloud is
not significantly evaporated ($h\nu_{\mathit{rf}} \gg \Delta
E+k_BT$, where T is the initial temperature of the cloud). From
the above considerations it becomes clear that for a given
temperature this method is only applicable for offset fields
$B_0\gg 3k_BT/\mu_B$.

For smaller magnetic offset fields, we use a second method (ii).
The initially polarized cloud evolves in the magnetic trap for a
certain time $t$. By applying a strong magnetic field gradient
after switching off the trap, the magnetically trapped atoms can
be spatially separated with respect to their substates before
imaging. This is equivalent to a Stern-Gerlach experiment. A
series of these images yields the time evolution of the population
of the trappable sub-levels. From the initial decay of the $m_S=3$
we obtain the loss rate $\widetilde{\beta}_{dr}$.
\begin{equation}\label{eq:decaylowoffset}
    \frac{dN_3(t)}{dt}=-\gamma_{bg}
    N_3(t)-\widetilde{\beta}_{dr}\frac{N_3^2(t)}{\overline{V}(t)}.
\end{equation}
It is not possible to distinguish between single (one atom lost
from $m_S=3$)and double (two atoms lost from $m_S=3$) spin flip
transitions with this method. Hence the relation between
$\widetilde{\beta}_{dr}$ and the cross sections changes to
\begin{equation}\label{eq:betalowoffset}
    \widetilde{\beta}_{dr}=\langle(\sigma_1+2\cdot\sigma_2)v_{rel}\rangle_{_{therm}}.
\end{equation}
The practicability of this method is limited by the magnitude of
the magnetic field gradient which can be applied to separate the
expanding thermal clouds before imaging.

The third method (iii) measures the heating of the atomic cloud
due to the energy released by dipolar relaxation processes. The
evolution of the temperature is given by
\begin{equation}\label{eq:heating}
 \frac{dT(t)}{dt}=\widetilde{\beta}_{dr}
 \frac{N_3(t)}{\overline{V}(t)}\Delta T.
\end{equation}
Since a two spin flip transition yields twice the energy of a
single spin flip transition the obtained value is
$\widetilde{\beta}_{dr}$ as in method (ii)
(Eq.\ref{eq:betalowoffset}). This method requires that the cloud
is in thermal equilibrium, i.e. that rethermalization by elastic
collisions is faster than the dipolar relaxation process itself.
%
%
As the velocity of atoms which have undergone a dipolar relaxation
event and the volume which is occupied by the collisional products
increases with increasing offset field, the elastic collision rate
drops and impairs the measured result.

\subsection{Preparation of the sample and experimental methods}

The basic setup used in our experiments has been described
previously \cite{Schmidt20031,Schmidt20033} and is only briefly
summarized in the following. Zeeman-slowed $^{52}$Cr atoms are
magneto-optically trapped in the magnetic field configuration of a
Ioffe--Pritchard trap in the cloverleaf configuration using the
$^7$S$_3$-$^7$P$_4$ transition. Since this transition is not
closed, we populate the metastable $^5$D$_4$ state via the weak
$^7$P$_4$-$^5$D$_4$ intercombination line. In this state the atoms
are decoupled from the trapping light and can be magnetically
trapped  because of the high magnetic moment ($6\mu_B$). We load
about $2\cdot10^8$ atoms continuously in the Ioffe--Pritchard type
trap (''CLIP''--trap \cite{Schmidt20032}). After switching off the
trapping light, we optically transfer (using light at 658\,nm) the
atoms back to the $^7$S$_3$ ground state, which possesses the same
magnetic moment($\mu=6\mu_B$). Subsequently, we compress the
sample and perform Doppler cooling in the magnetic trap. Using
this method, we obtain a nearly polarized ($m_S=3$) atomic
ensemble with an initial density of about $10^{11}$\,atoms/cm$^3$
and a temperature of $275\pm25\,\mu$K.

This is the starting point for most of our measurements of dipolar
relaxation at high magnetic offset fields ($17-44$\,G) using
method (i). However, for smaller offset fields ($<$25\,G), it was
necessary to reduce the temperature further by rf-evaporation to
realize a cut-off parameter for the rf-shield of $\eta\geq5$
\cite{Masuhara1988}. After preparation, the rf-shield is applied
and the atomic ensemble is allowed to evolve in the trap for a
variable time $t$. The rf-frequency is set to $1$\,MHz below
$7\Delta E/6h$. Subsequently, the shadow of the atoms is imaged
onto a CCD-camera by illuminating the atoms for $100\,\mu$s with a
circularly polarized beam after $5$\,ms of free ballistic
expansion in a homogenous magnetic field.

The number of remaining atoms $N_3(t)$ and the mean volume
$\overline{V}(t)$ are extracted from a 2D-Gauss fit to absorption
images using the separately measured trap frequencies (at $27$\,G
offset field: $\nu_{radial}=120$\,Hz, $\nu_{axial}=73$\,Hz). A
series of these measurements with different evolution times $t$
yields $N_3(t)$. The temperature $T(t)$ is obtained by recording
the ballistic expansion of the atom cloud at different evolution
times.

For measurements at low offset fields ($<$2\,G), we use the second
approach (ii). Since the temperature obtained after Doppler
cooling is too high to separate the atoms with the available field
gradient, we reduce the temperature of the sample to $50\,\mu$K by
rf-evaporative cooling at an offset field of 0.7\,G (trap
frequencies for this configuration are typically
$\nu_{radial}=806$\,Hz, $\nu_{axial}=42$\,Hz). Subsequently, we
turn off the rf-frequency and change the offset field for a free
evolution time $t$ to the desired value. Shortly (10\,ms) before
the Stern-Gerlach sequence, the field is changed back to 0.7\,G.

\begin{figure}
\begin{center}
\includegraphics[width=0.9\columnwidth]{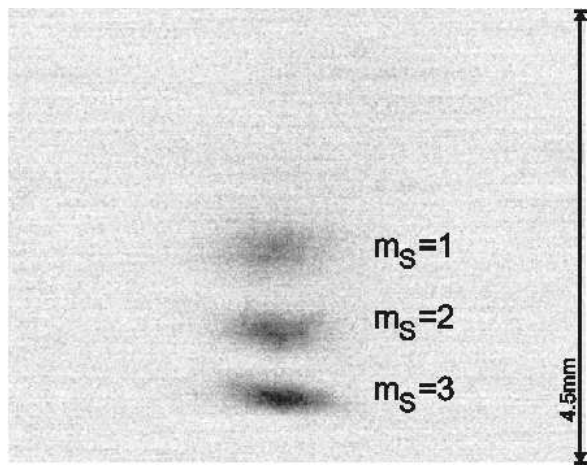}
\end{center}
\caption{Typical fluorescence picture taken after separation of
the sub-levels in a Stern-Gerlach type experiment and $5$\,ms time
of flight. The sub-levels have been accelerated downwards.
}\label{fig:sterngerlachpic}
\end{figure}

The separation of the substates is accomplished by switching off
the axially confining pinch and bias fields of our clover-leaf
trap and simultaneously applying a homogenous field in the imaging
direction which is perpendicular to the offset field direction and
diagonal to the leaf position of our trap. By this means the
center of the trap is displaced radially and the atoms get
accelerated in the gradient field of the leaves ($170$\,G/cm)
along an axis perpendicular to the imaging direction. After a
$1.7$\,ms acceleration phase, all magnetic fields are turned off
and the atoms expand freely for another $5$\,ms before the
spatially separated clouds are imaged on the CCD camera by
illuminating them for $1$\,ms with a calibrated light field. The
data analysis is perform by averaging three pictures taken under
identical experimental conditions for each evolution time. An
example for such an image taken at $0.3$\,G is depicted in Fig.
\ref{fig:sterngerlachpic}.
The clouds have been accelerated downwards. The cloud containing
atoms in the energetically highest state ($m_S=3$) therefore
appears at the lowest position.

The number of atoms in each substate are obtained by 2D-Gauss fits
within an equally sized rectangular area around each cloud.
Because of the long exposure time and the trap geometry being
changed while switching off the trap, temperature and volume were
obtained in separate measurements where we record the free
expansion of the whole cloud at different evolution times with a
exposure time of $200\,\mu$s.

We determine the offset field by probing the minimum of the
trapping field via rf removal of trapped atoms. The achieved
accuracy of the values is about 300\,mG for measurements at high
offset values and \linebreak 100\,mG for the measurements at low
offset fields. The background gas collision rate
$\gamma_{bg}=1/200$\,s is obtained from a separate measurement
with a low density atomic sample. This reduces the fitting
parameters to the initial number of atoms and $\beta_{dr}$. The
increasing volume due to heating during the measurement was
included in the fit. The errors in the determination of the number
of atoms is about 30\%. The error bars shown in the figures are
the square root of the diagonal elements of the covariance matrix
for the fitting parameters obtained from a least-squares fit.

\subsection{Results}
\begin{figure}
\begin{center}
\includegraphics[width=0.9\columnwidth]{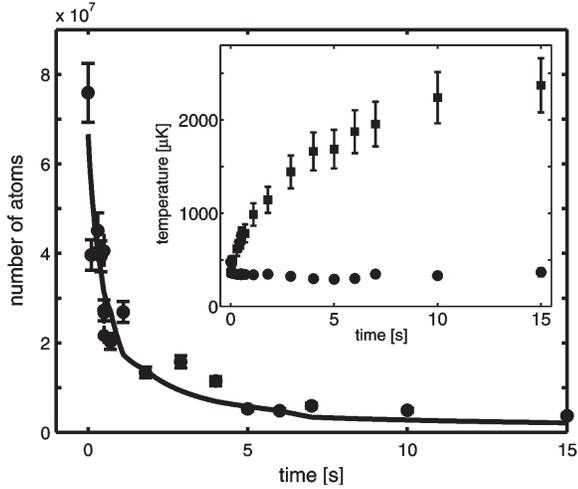}
\end{center}
\caption{Decay measurement using method (i). The number of atoms
$N_3$(t) remaining in the $m_S=3$ state after a variable free
evolution time in the trap at an offset field of $27$\,G is
depicted.
The curve represents a fit using the solution of Eq.
\ref{eq:decayhighoffset}. The inset provides the temperature
evolution with (bullets) and without (squares)
rf-shield.}\label{fig:decayhighoffset}
\end{figure}

We have performed measurements using the first method (i) for
different magnetic offset fields ranging from $17$ to $52$\,G. An
example of a typical decay curve of the population $N_3(t)$ of the
$m_S=3$ level, taken at an offset field of $27$\,G, is plotted
in Fig. \ref{fig:decayhighoffset}.
The fit to a two body decay using the solution of Eq.
(\ref{eq:decayhighoffset}) yields $\beta_{dr}=(2.5\,\pm\,0.5)
\cdot10^{11}\,cm^3/s$ and is indicated by the black curve in the
figure. The inset to this figure demonstrates the evolution of the
temperature with (bullets) and without (squares) rf-shield.
Hereby, the temperature is derived from the width of the cloud.
Without rf-shield, strong heating with an initial rate of more
than 600\,$\mu$K/s is observed. With rf-shield, the temperature
remains constant.

\begin{figure}
\begin{center}
\includegraphics[width=0.9\columnwidth]{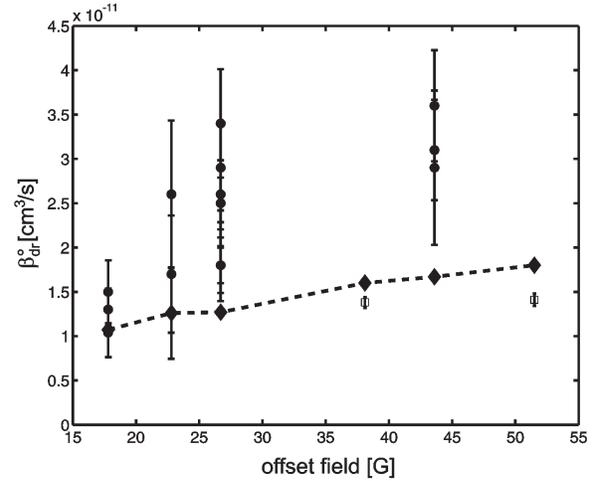}
\end{center}
\caption{Dependence of the dipolar rate constant
$\beta^\circ_{dr}$ on the offset field $B_0$. Depicted are
$\beta^\circ_{dr}$ (bullets), the calculated values (diamonds,
connected by straight lines to guide the eye) and values obtain
from a different setup (open squares). The calculated values are
obtained by using Eq. (\ref{eq:betahighoffset}) and the
experimental parameters $T$ and $B_0$}.\label{fig:beathighoffset}
\end{figure}

The results of the decay measurements are summarized in Fig.
\ref{fig:beathighoffset}. The scatter of the data for a given
offset field indicates the uncertainty in our measurement. As
expected, the loss coefficient increases with increasing field.
Each bullet represents a separate decay measurement. Also included
are two values (depicted by open squares), which were measured in
a different experimental setup using the same method. The diamonds
which are linked by a straight lines to guide the eye, represent
the results obtained from the theoretical model using Eq.
(\ref{eq:betahighoffset}) and the experimental parameters $T$ and
$B_0$. Though there is a slight discrepancy between the data and
the model calculation, the experimental values are reproduced
within a factor of two.
The deviation between the two experimental data sets is attributed
to systematic errors in the determination of the density in these
experiments.

\begin{figure}
\begin{center}
\includegraphics[width=0.9\columnwidth]{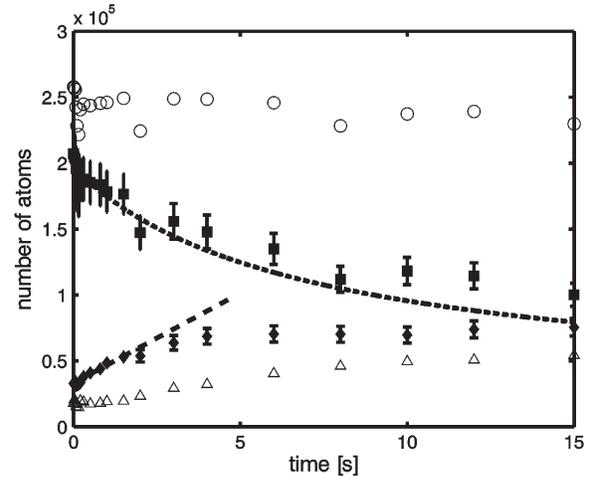}
\end{center}
 \caption{Redistribution among the sub-levels. While the total number of atoms ($N_{tot}$)(circles) stays constant, the population of the initially almost polarized sample gets redistributed by dipolar relaxation and spin exchange collisions. The population $N_{mS}$ of each sub-levels $m_S=1,2$ and $3$ is indicated by triangles, diamonds and squares, respectively. Also shown are the fitting curves to the decay of the $m_S=3$ (dotted line) and loading of the $m_S=2$ state (dashed line).}\label{fig:decaylowoffset}
\end{figure}

Measurements of the dipolar relaxation rate at low offset fields
have been performed using method (ii). Fig.
\ref{fig:decaylowoffset} shows a typical result of a Stern-Gerlach
Experiment at an offset field of $0.7$\,G. The figure contains the
evolution of the number of atoms in each sub-level and the total
number of atoms ($N_{tot}(t)=N_1(t)+N_2(t)+N_3$(t)) during the
first $15$\,s. While the total number of trapped atoms (circles)
stays approximately constant, about 35\% of the atoms initially in
the $m_S=3$ state (squares) get redistributed among the other
trapped states.
The initially prepared atomic cloud is not fully polarized even at
$t=0$\,s. We observe a small fraction of the cloud (about 20\%) in
the $m_S=1,2$ states (triangles, diamonds). The origin of this
population remains unclear and might either be caused by
incomplete rf evaporation or by Majorana spin flips
\cite{Petrich1995} during switching of the magnetic fields. An
indication that evaporative cooling populates these states is the
slightly decreasing number of atoms which can be observed in the
first few hundred ms in the $m_S=1,2$ states in samples with lower
densities.

\begin{figure}
\begin{center}
\includegraphics[width=0.9\columnwidth]{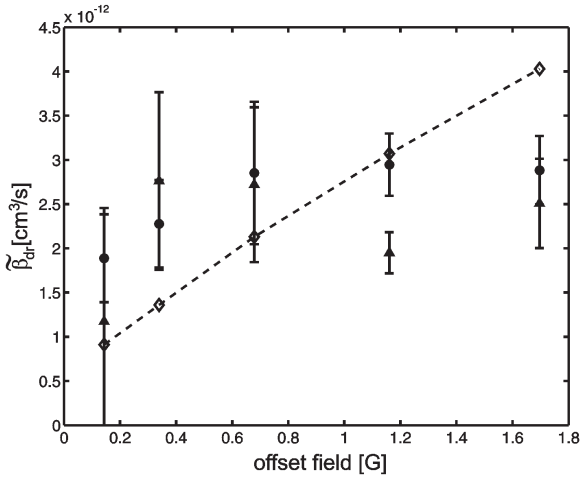}
\end{center}
    \caption{Dependence of the dipolar relaxation rate $\widetilde{\beta}_{dr}$ on the offset field $B_0$ at low magnetic fields. The experimental values obtained by method (ii) are indicated by bullets. The diamonds connected by straight lines to guide the eye are calculated using Eq. (\ref{eq:betalowoffset}) and experimental parameters $B_0$ and T. Also included are the values (triangles) obtained from the temperature increase of the sample.}\label{fig:beatlowoffset}
\end{figure}

The decay of the $m_S=3$ population is non-exponen\-tial. The
dynamics is described by two coupled, non-linear differential Eqs.
(\ref{eq:decaylowoffset}) and (\ref{eq:heating}) (where
$\overline{V}(t)\propto \linebreak T^{3/2}(t)$). By assuming a
linear increase in the volume which is describing by our data
well, we obtain a solution for Eq. (\ref{eq:decaylowoffset}). A
fit to data for which the population of the $m_S=2$ state is below
$25\%$ yields a dipolar relaxation rate constant of
$\widetilde{\beta}_{dr}=(3.1\,\pm\,0.4)\cdot10^{-12}$\,cm$^3$s$^{-1}$.
The fit to N$_3$(t) is illustrated by a dotted curve. We repeated
this measurement for 5 different offset fields ranging from $0.15$
to $1.7$\,G. The dependence of the measured values on the offset
field is shown in Fig. \ref{fig:beatlowoffset}. Again, the
diamonds which are connected by straight lines to guide the eye
are obtained from Eq. (\ref{eq:betalowoffset}) using the measured
experimental parameters $B_0$ and $T$ with no other adjustable
parameter.
The agreement of the calculated values with the experimental data
is reasonable, although the square root dependence of the decay
rate on the magnetic field is not exactly reproduced.

We have also investigated the loading process of the $m_S=2$ state
in order to deduce the two body-loss coefficient $\beta_{2}$ of
this state. For short lifetimes, the loading rate into the $m_S=2$
state can be exclusively attributed to the loss out of the $m_S=3$
state. Neglecting background gas collisions, we obtain the
following rate equation:
\begin{equation}\label{eq:mj2loading}
    \frac{dN_2(t)}{dt}=-\frac{dN_3(t)}{dt}-\beta_{2}\frac{N_2^2(t)}{\overline{V}_2(t)}
\end{equation}
where $\overline{V}_2=(3/2)^{3/2}\overline{V}_3$ is the mean
volume of the thermalized $m_S=2$ cloud. As spin exchange
collisions and dipolar relaxation are possible in this state,
$\beta_2$ contains both processes. Assuming a constant volume, the
equation can be solved. A fit of the first order power series
expansion of the solution to our data yields a value of
$\beta_2=(1.1\,\pm\,0.2)\,\cdot\,10^{-10}\,cm^3s^{-1}$, where we
have taken the average of the measurements between $0.1$ and
$1.7$\,G. The fit is indicated in Fig. \ref{fig:decaylowoffset} by
a dashed curve. The obtained values are shown in the inset of Fig.
\ref{fig:beatlowoffset}. The values of the coefficients are about
a factor 50 higher than the dipolar relaxation coefficient of the
$m_S=3$ state. This can probably be attributed to the additional
decay channels mentioned in Sec.\ref{sec:theory}.

We observe that the population in the $m_S=1$ state increases
initially much slower than the population in the $m_S=2$ state.
This indicates that $m_S=3$ state does not directly decay to the
$m_S=1$ state via dipoar relaxation. The almost constant total
number of atoms gives further evidence that the observed loss from
the $m_S=3$ state involves mainly spin-flips with $\Delta m_S=1$.
Thus, the first order Born approximation is a valid approach to
describe dipolar relaxation in our system.

\begin{figure}
\begin{center}
\includegraphics[width=0.9\columnwidth]{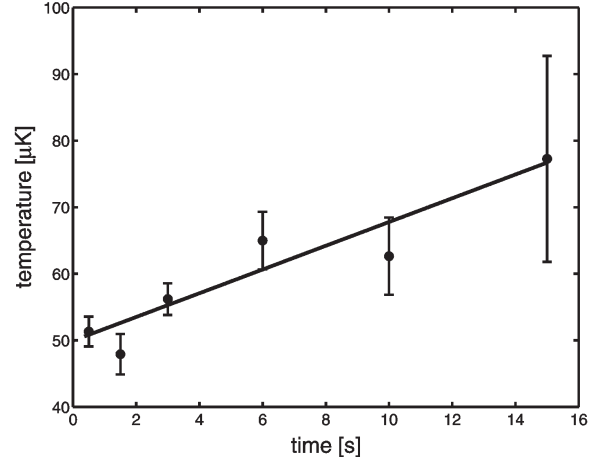}
\end{center}
    \caption{Heating due to dipolar relaxation at a low offset field (0.7\,G). The solid line is a linear fit to the data.}\label{fig:Tincrease}
\end{figure}

An example for the time evolution of the mean temperature is
depicted in Fig. \ref{fig:Tincrease}. Even at a rather low offset
field of $0.7$\,G the temperature of initially at $50$\,$\mu$K
increases rapidly with an approximate rate of $3\,\mu$K\,s$^{-1}$.
We estimate the dipolar relaxation rate constant using method
(iii) by solving Eq. (\ref{eq:heating}) using the solution for the
number of atoms in the $m_S=3$ state obtained above. The results
of a fit to the linearized solution are included in Fig.
\ref{fig:beatlowoffset} (triangles) and are in agreement with the
results using method (ii).

\begin{figure}
\begin{center}
\includegraphics[width=0.9\columnwidth]{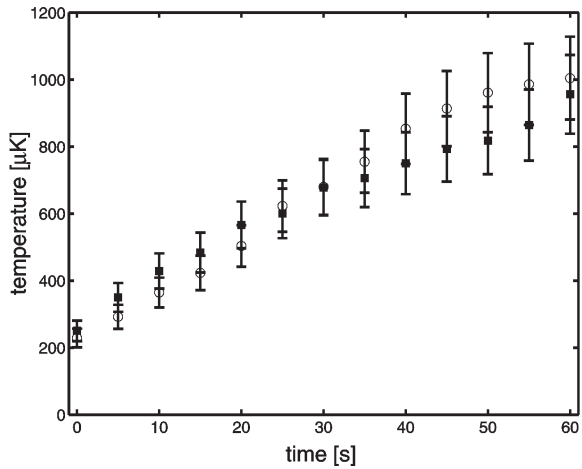}
\end{center}
\caption{Heating rate due to dipolar relaxation at $20$\,G offset
field for the two isotopes $^{52}$Cr (open circles) and $^{50}$Cr
(squares). The initial density for both isotopes is roughly
$n_0=10^{10}$\,cm$^{-3}$.}\label{fig:isotopes}
\end{figure}

The dipolar relaxation rate is calculated supposing that the
inelastic process is dominated by the dipole-dipole interaction.
Details of the molecular interaction are not taken into account.
We have tested this assumption by comparing the heating of the
atomic cloud due to dipolar relaxation for the two isotopes
$^{52}$Cr and $^{50}$Cr. The temperature evolution of the atomic
clouds at an offset field of $B_0=20$\,G is shown in Fig.
\ref{fig:isotopes}.
Here, the temperature was derived from the $1/\sqrt{e}$ width
$\sigma_y$ of the cloud. Equally high heating rates
($13$\,$\mu$Ks$^{-1}$) yielding the same loss coefficient can be
observed for both clouds prepared with the same initial atom
number density of $10^{10}$\,atoms/cm$^3$. The data shown in Fig.
\ref{fig:isotopes} indicate that the loss due to dipolar
relaxation is independent of the details of the interatomic
potential. Since the scattering length for $a(^{50}Cr)=50\,a_0$
and $a(^{52}Cr)=170\,a_0$ differs significantly
\cite{Schmidt20033} one would expect that any effect due to the
'molecular' interaction also differs for the two isotopes.

\section{Conclusion}
\label{sec:conclusion}

We have presented three different methods to obtain the dipolar
relaxation rate constant in magnetically trapped clouds of
chromium atoms. Using these methods we have measured the
dependence of the rate constant on the magnetic offset field of
the trapping potential. We find a typical dipolar relaxation rate
constants of $\widetilde{\beta}_{dr}=4\cdot
10^{-12}$cm$^3$s$^{-1}$ at a magnetic offset field of B$_0=1$\,G.
Thus, even at low magnetic fields, the loss coefficient for
chromium is one order of magnitude larger than the one reported
for Cs in lowest hyperfine level (F=3) \cite{Guery-Odelin1998} and
another 3 orders of magnitude larger than in Na
\cite{Goerlitz2003}.

A theoretical estimation of the inelastic cross sections for
spin-flip transitions due to the magnetic dipole-dipole
interaction, predicts cross sections for single and double spin
flip transition which scale with the total angular momentum to the
third and second power, respectively. The values predicted by the
calculation are in good agreement with the experimentally observed
values. The influence of the molecular potential on the dipolar
relaxation cross section for chromium was found to be
insignificant.

Our findings could be especially important for experiments which
aim at producing high density samples of atoms and molecules with
high magnetic or electric dipole moments in magnetostatic or
electrostatic traps
\cite{Weinstein19981,Weinstein19982,Bethlem2000}. Large inelastic
two-body collision rates are to be expected with increasing
dipole-dipole interaction, if atoms are not trapped in the lowest
energy state of the dipole.

As has already been pointed out in \cite{Guery-Odelin1998} dipolar
relaxation can be almost totally suppressed by polarizing the
sample in the energetic lowest spin state, which requires
different trapping schemes. In this state neither dipolar
relaxation nor spin exchange collisions are possible at reasonable
offset field. The only expected loss mechanism in this case is
three-body recombination. However, because of the long-range
nature of the dipole potential the latter could also be increased
compared to atoms with a small dipole moment. As a quite general
conclusion, strong dipolar gases can only be cooled to degeneracy
in traps that keep the dipole moment in its lowest energy state.

Our results, though not promising at first glance, will help us to
develop a strategy for reaching Bose-Einstein condensation even
under the special conditions set by the large magnetic moment of
chromium. In cesium, losses due to inelastic collisions in a
magnetic trap have prevented Bose-Einstein condensation for a long
time in a similar way as in the chromium case. However, a careful
study of the nature of the inelastic losses has finally allowed to
devise a successful strategy for condensation \cite{Weber2003}.

This work was supported by the Deutsche Forschungsgemeinschaft and
the RTN network 'Cold Quantum \linebreak Gases'.

\end{document}